\documentclass[%
12pt,
reprint,
 superscriptaddress,
 amsmath,amssymb,
 aps,UTF8,
]{revtex4-1}

\usepackage{graphicx}% Include figure files
\usepackage{dcolumn}% Align table columns on decimal point
\usepackage{bm}% bold math
\usepackage{xcolor}
\usepackage{url}
\usepackage{float}
\usepackage{tabularx}
\usepackage{lipsum}
\usepackage{array}

\usepackage{booktabs}
\usepackage{subfigure}
\usepackage{multirow}

\begin{document}

%\preprint{APS/123-QED}

\title{Unraveling Proton Strangeness: Determination of the Strangeness Sigma Term with Statistical Significance}%% Force line breaks with \\
%%\thanks{A footnote to the article title}%
\author{Wei Kou}
\email{kouwei@impcas.ac.cn}
\affiliation{Institute of Modern Physics, Chinese Academy of Sciences, Lanzhou 730000, China}
\affiliation{School of Nuclear Science and Technology, University of Chinese Academy of Sciences, Beijing 100049, China}
%
%\author{Bingang Guo}
%\affiliation{Institute of Modern Physics, Chinese Academy of Sciences, Lanzhou 730000, China}
%\affiliation{School of Nuclear Science and Technology, University of Chinese Academy of Sciences, Beijing 100049, China}

%
%
\author{Xurong Chen}
\email{xchen@impcas.ac.cn (Corresponding author)}
\affiliation{Institute of Modern Physics, Chinese Academy of Sciences, Lanzhou 730000, China}
\affiliation{School of Nuclear Science and Technology, University of Chinese Academy of Sciences, Beijing 100049, China}



\begin{abstract}
This study delves into the contribution of the strange quark within the proton, which influences several fundamental proton properties. By establishing a robust relationship between the proton's quantum anomalous energy and the sigma term, we successfully extract the contribution of the strangeness sigma term in the proton, obtaining it from parameterized fits of the differential cross-section using exponential and dipole forms. The extracted sigma term values are estimated to be $455.62\pm64.60$ MeV (exponential fit) and $337.74\pm106.79$ MeV (dipole fit).
Additionally, we present novel results for the proton trace anomalous energy, observing values of $0.13\pm0.02$ (exponential fit) and $0.15 \pm 0.03$ (dipole fit). Our analysis integrates the novel data from the $J/\psi-007$ and GlueX collaborations, to some extent providing an experimental phenomenological window into the non-zero strange quark content inside proton, with the statistical significances: $7.1\sigma$ (exponential fit) and $3.2\sigma$ (dipole fit), respectively. Furthermore, we discussed the possibility of scheme-independence in the extraction results and examine the uniformity of sigma terms obtained from the datasets provided by the two collaborations. Our analysis may reveals compatibility between the extracted results of the respective experiments.

\end{abstract}

\pacs{24.85.+p, 13.60.Hb, 13.85.Qk}% PACS, the Physics and Astronomy
%% \keywords{Suggested keywords}%Use showkeys class option if keyword
\maketitle

%\tableofcontents

 \section{Introduction}
 \label{sec:Intro}
The parton model provides insight into the dynamics of the internal components of the proton. Protons, as complex bound-state particles comprising quarks and gluons, have been the subject of extensive discussion regarding their external properties, such as spin, mass, charge radius, and other attributes \cite{Ji:2020ena,Ji:2021mtz,Gao:2021sml}. Understanding how the contributions of individual components of the proton manifest in its external properties is a complex and ongoing issue. Notably, experimental evidence for the contribution of partons other than up and down quarks remains insufficient. The role of strange and charm quarks in protons continues to be a topic of significant interest in the field \cite{Ball:2022qks,An:2005cj,Zou:2005xy,Zou:2006tw,Sapozhnikov:2011zza}. Studies investigating the strangeness contribution to proton magnetic moment \cite{Zou:2005xy} have provided valuable insights into the intrinsic strange quark within the proton.

The contribution of quarks to the proton mass can be summarized by the quark scalar matrix elements -- sigma terms. They are defined as 
\begin{equation}
	\sigma_{qN}=\langle N\mid m_q\bar{q}q\mid N\rangle,
	\label{eq:sigma term def}
\end{equation}
where $|N\rangle$ is proton state in the rest frame, $q(\bar{q})$ the quark field operators, $m_q$ the current quark mass of flavor $q$. For the u- and d-flavor quark cases, the $\pi N$ sigma term is introduced and denoted by $\sigma_{\pi N} = m_l\langle N|u\bar{u}+d\bar{d}|N\rangle $ with the average mass of $u$, $d$ quarks $m_l$. The sigma terms of the nucleon are the main source of uncertainty in the calculation of the scattering for spin-independent nucleon with dark matter \cite{Ellis:2018dmb}, so it is particularly important to measure them precisely from experiments. Previous studies of the nucleon sigma terms can be found in \cite{Gasser:1990ce,Pavan:2001wz,Ohki:2008ge,Toussaint:2009pz,Takeda:2010cw,Alarcon:2011zs,Alarcon:2012nr,BMW:2011sbi,QCDSF-UKQCD:2011qop,QCDSF:2011mup,Sapozhnikov:2011zza,Shanahan:2012wh,Ohki:2012jyg,XQCD:2013odc,Alvarez-Ruso:2013fza,Junnarkar:2013ac,An:2014aea,Lutz:2014oxa,Ren:2014vea,Yang:2015uis,Hoferichter:2015dsa,Abdel-Rehim:2016won,RuizdeElvira:2017stg,Varnhorst:2018hrk,Alexandrou:2019brg}.  A previous determination from high-precision data gives the value of the $\sigma_{\pi N}$ as $59.1\pm 3.5$ MeV \cite{Hoferichter:2015dsa}. The current discussion on the experimental perspective of the contribution of strange quarks $\sigma_{sN}$ to the scalar charge is still insufficient. Fortunately there are many lattice QCD calculations \cite{Yang:2015uis,Alexandrou:2019brg} as well as results from chiral perturbation theory performances \cite{QCDSF:2011mup,Alarcon:2012nr} available to us. In addition, we note that there is the work that suggests constraining the strangeness component by measuring the mass shift of $\phi$ meson in nuclear matter \cite{Gubler:2014pta}. At this point we may consider that the strangeness component of the proton mass can be linked to its corresponding scalar charge or sigma term.

The sigma term is related to the QCD gauge invariant Hamiltonian decomposition, which comes from Ji's work \cite{Ji:1994av,Ji:1995sv,Ji:2021mtz,Ji:2021pys}. The decomposition of QCD Hamiltonian is written as
\begin{equation}
	H_{QCD}=H_q+H_g+H_m+H_a.
	\label{eq:hamiltionian}
\end{equation}
They represent, in order, the quark energy, the gluon field energy, the quark mass contribution, and the quantum  anomalous energy with proton mass $M_N=\langle N|H|N\rangle/\langle N|N \rangle$ in the rest frame. 
\begin{equation}
	\begin{gathered}
		\begin{aligned}
			M_q=\frac{3}{4}\left(a-\frac{b}{1+\gamma_m}\right)M_N,\ \ \ &M_g=\frac{3}{4}(1-a)M_N,
			\end{aligned} \\
		M_m=\frac{4+\gamma_m}{4\left(1+\gamma_m\right)}bM_N,\quad M_a=\frac{1}{4}(1-b)M_N,
\end{gathered}
\label{eq:mass dec}
\end{equation}
where $\gamma_m$ is the anomalous dimension of quark mass \cite{Buras:1979yt}. These four terms represent the eigenvalues of the mass corresponding to the four parts after the decomposition of the Hamiltonian. The parameters $a(\mu)$ is the momentum fraction carried by quarks in the infinite momentum frame, and $b$ is related to the scalar charge of the proton as $b=\langle N|m_u\bar{u}u+m_d\bar{d}d+m_s\bar{s}s|N\rangle/M_N$.  More specifically it can be expressed as
\begin{equation}
	bM_N=\sigma_{\pi N}+\sigma_{sN}.
	\label{eq:bmn}
\end{equation}

According to the previously mentioned definition of the parameter $b$, it is determined by the scalar charge from $u$, $d$, and $s$ quarks. While precise experimental measurements of $\sigma_{\pi N}$ are already available, as mentioned earlier, the direct experimental determination of the strangeness component presents challenges \cite{Gao:2015aax}. In addition to using measurements of the mass shift of the $\phi$ meson in nuclear matter to determine the strangeness sigma term \cite{Gubler:2014pta}, there is another possible way to indirectly obtain the strangeness sigma term experimentally by relating it to the parameter $b$ in Eq. (\ref{eq:bmn}) \cite{Wang:2019mza, Kou:2021bez, Duran:2022xag, Wang:2022tzw}. The latter approach is related to the near-threshold photo-production process of the ground-state charmonium \cite{Kharzeev:1995ij,Kharzeev:1998bz}. Fortunately, experimental measurements of the process \cite{Duran:2022xag,GlueX:2019mkq,GlueX:2023pev} have become substantial in recent years. We use recent data for theoretical analysis and provide the strangeness contribution to the proton sigma term. The discussion of details is presented in the following sections.

\smallskip

\section{Extraction methods and experiments}
\label{sec:method}
%\noindent\emph{2.$\;$Extraction methods and experiments} ---
The near-threshold photo-production process of $J/\psi$ is commonly used to extract the Gravitational Form factors (GFFs) of the proton, which are related to the QCD Energy-Momentum-Tensor. The GFFs of the proton can be utilized to reveal the mass distribution as well as mechanical properties such as pressure inside the proton, as demonstrated by Holographic QCD \cite{Mamo:2019mka} or the Generalized Parton Distribution functions (GPDs) \cite{Guo:2021ibg}. These concepts should essentially provide information about GFFs.

In Refs. \cite{Barger:1975ng,Sivers:1975ds,Babcock:1977fi,Kharzeev:1998bz}, the calculation of the differential cross section of $J/\psi$ photo-production at near-threshold via momentum transfer $t$ is presented as
\begin{equation}\begin{aligned}
		&\left.\frac{d\sigma_{\gamma N\to J/\psi N}}{dt}\right|_{t=0} \\
		&\left.=\frac{3\Gamma(J/\psi\to e^+e^-)}{\alpha m_{J/\psi}}\left(\frac{k_{J/\psi N}}{k_{\gamma N}}\right)^2\frac{d\sigma_{J/\psi N\to J/\psi N}}{dt}\right|_{t=0},
		\label{eq:jpsi xsection}
\end{aligned}\end{equation}
where $\begin{aligned}k_{ab}^2=[s-(m_a+m_b)^2][s-(m_a-m_b)^2]/4s\end{aligned}$ denotes the momentum square of the reaction in the center-of-mass frame with photon-proton center-of-mass energy $\sqrt{s}$, $\alpha=1/137$ is the fine structure constant and $\Gamma$ stands for the partial $J/\psi$ decay width. Using the vector meson dominance (VMD) approach, the differential cross section of $J/\psi-N$ at low energy is expressed with the $J/\psi-N$ elastic amplitude $F_{J/\psi N}$ as
\begin{equation}
	\left. \frac{d\sigma_{J/\psi N\to J/\psi N}}{dt}\right|_{t=0}=\frac1{64\pi}\frac1{m_{J/\psi}^2(\lambda^2-M_N^2)}F_{J/\psi N}^2,
	\label{eq:anmplitude}
\end{equation}
where $\lambda=(pK/m_{J/\psi})$ is the nucleon energy in the charmonium rest frame, and $p$, $K$, are the four-momenta of the target nucleon, $J/\psi$, respectively.

The $J/\psi-N$ elastic scattering amplitude $F_{J/\psi N}$, a key component of this study, is determined using the heavy quark non-relativistic approximation \cite{Kharzeev:1995ij}. In this context, the ground state $J/\psi$ system is considered to be governed by Coulomb-like interactions between positive and negative charm quarks. Notably, the Bohr radius of the $J/\psi$ meson, given by $r_0=4/(3\alpha_s m_c)$, depends on the charm quark mass $m_c$ and strong coupling $\alpha_s$. Following the expression in Ref. \cite{Kharzeev:1995ij}, the $J/\psi-N$ scattering amplitude is given by
\begin{equation}\begin{aligned}
		F_{{J/\psi}N}=r_0^3d_2\frac{8\pi^2}{27}(1-b)M_N^2m_{J/\psi},
		\label{eq:FXN}
\end{aligned}\end{equation}
with $bM_N$ in Eq. (\ref{eq:bmn}). Here, $d_n$ represents the Wilson coefficient defined in \cite{Peskin:1979va}. The right-hand side of the above equation is derived from the discussion surrounding Eq. (\ref{eq:bmn}). It is important to note that Eq. (\ref{eq:FXN}) exhibits dimensional inconsistencies with the formulas in Refs. \cite{Wang:2019mza,Wang:2022tzw,Kharzeev:1995ij}. Therefore, we adopt the correct dimensional results as presented in Ref. \cite{Kharzeev:1998bz}, consistent with our prior work \cite{Kou:2021bez}.

We conduct our analysis utilizing the recently published differential cross-section data by the $J/\psi-007$ and GlueX Collaborations at JLab \cite{Duran:2022xag,GlueX:2023pev}. As outlined in Eqs. (\ref{eq:jpsi xsection}-\ref{eq:FXN}), the forward differential cross-section for each experimental energy set is required for our analysis. In this communication, we opt for the exponential form $d\sigma/dt =d\sigma/dt|_{t\to0}\times \exp({-Bt})$ (exponential) and $d\sigma/dt=d\sigma/dt|_{t\to0}\times (1/(1-t/M_A^2))^4$ (dipole) \cite{Duran:2022xag} to fit the data.

Different photon energies that lead to $J/\psi$ production near the threshold are expected to result in distinct forward scattering cross-sections. Previous studies \cite{Wang:2019mza,Wang:2022tzw} relied on experimental data from \cite{GlueX:2019mkq} with only one energy point, making it challenging to determine whether the anomalous energy trace in protons was scale-dependent. However, recently published experimental data \cite{Duran:2022xag,GlueX:2023pev} and research on extracting proton mass radii and trace anomalous energy have revealed perplexing energy dependencies \cite{Duran:2022xag,Kou:2021bez}. Despite the expectation that these observations should not be scale-dependent, we aim to address these questions in the following sections.

\smallskip

\section{Numerical results and discussions}
\label{sec:results}
\subsection{Calculation and discussion}
\begin{figure*}[htbp]
	\centering  %图片全局居中
	\subfigure{
	\includegraphics[width=0.99\textwidth]{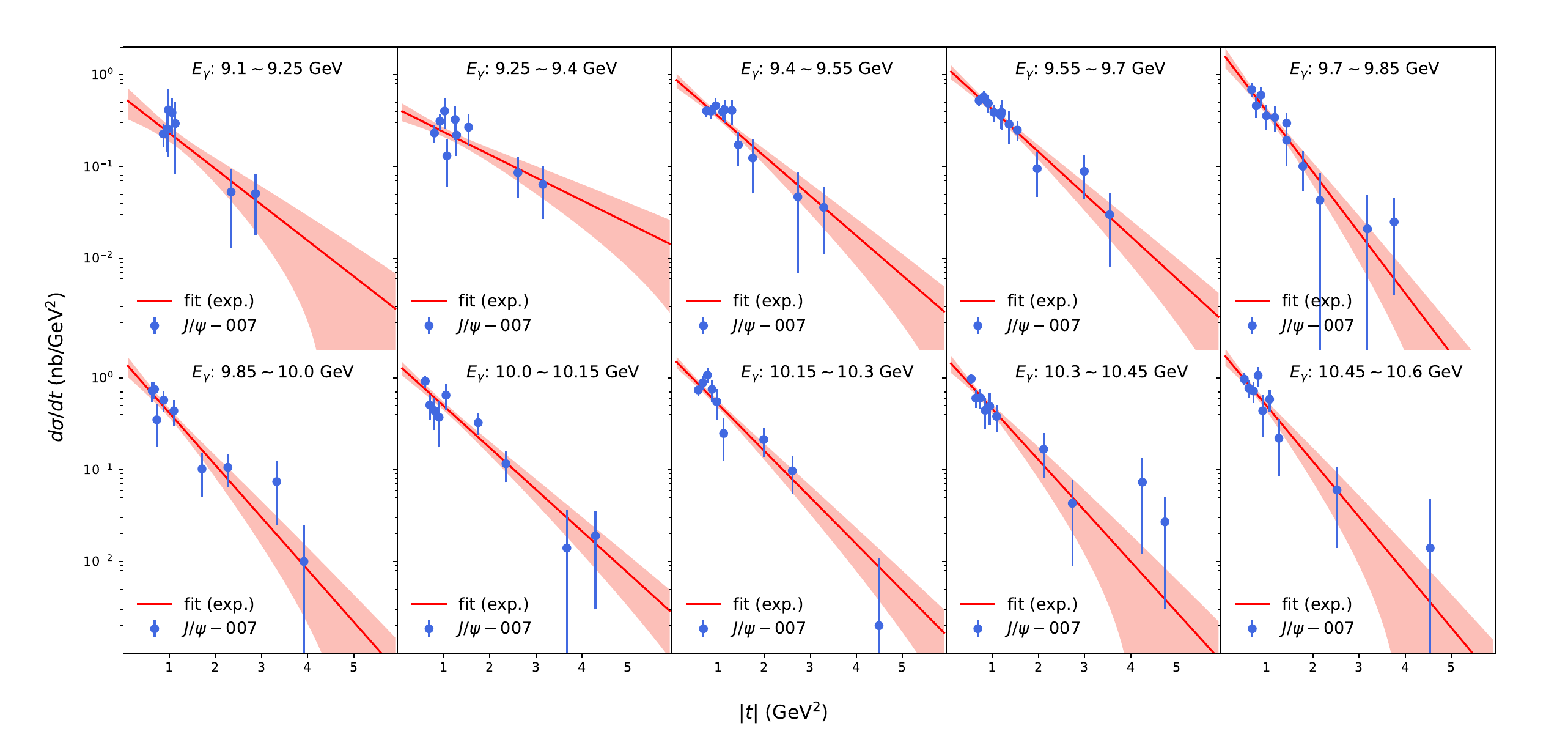}}
	\subfigure{
		\includegraphics[width=0.99\textwidth]{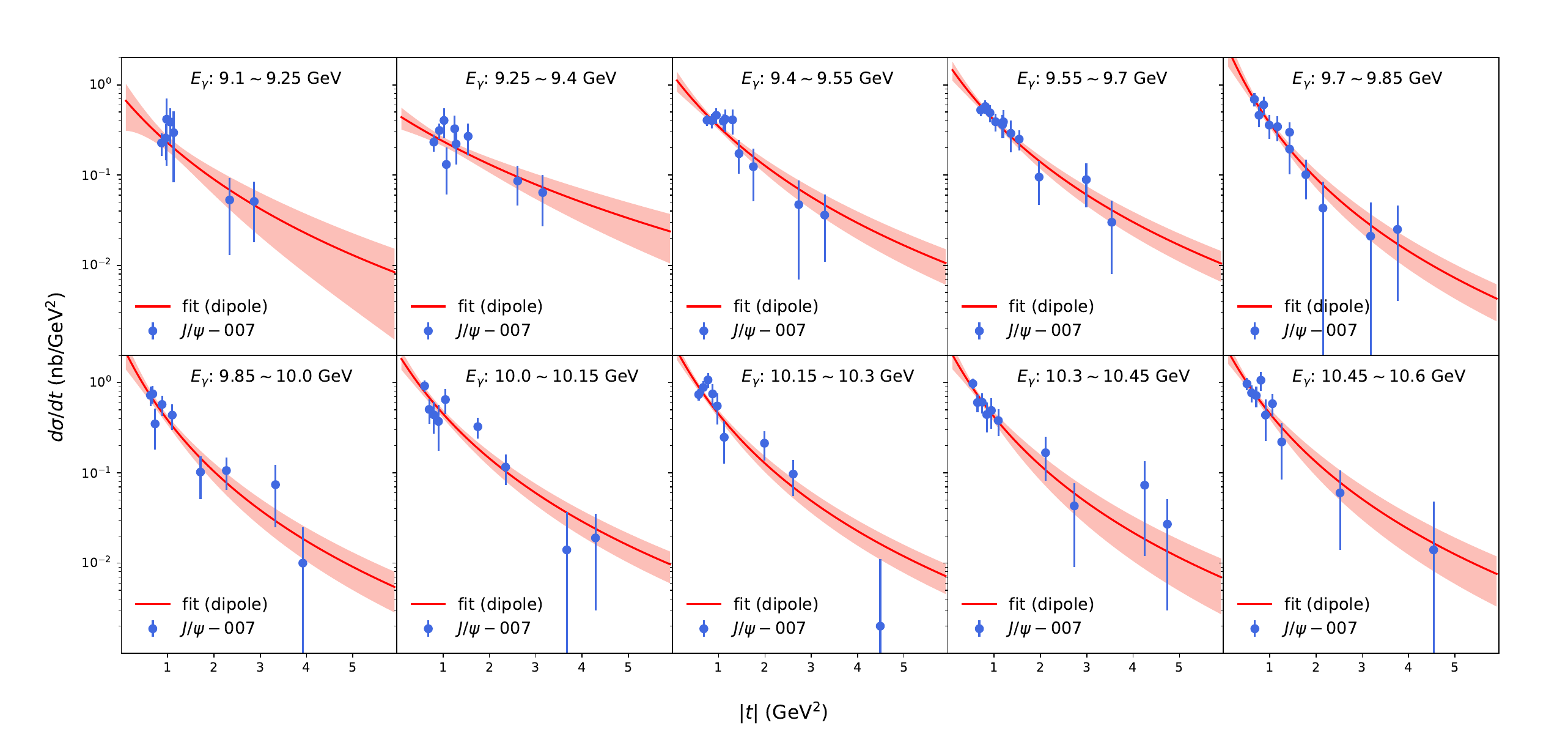}}
	\caption{Exponential (up) and dipole (down) fits to the differential cross section $d\sigma/dt$ using the data from $J/\psi-007$ collaboration \cite{Duran:2022xag}. The orange band shows its uncertainty. The parameters are listed in Tables \ref{tab:jpsi007} and \ref{tab:jpsi007-dipole}.}
	\label{fig:fit_hall_c}
\end{figure*}

\begin{figure*}[htbp]
	\centering  %图片全局居中
		\subfigure{
		\includegraphics[width=0.9\textwidth]{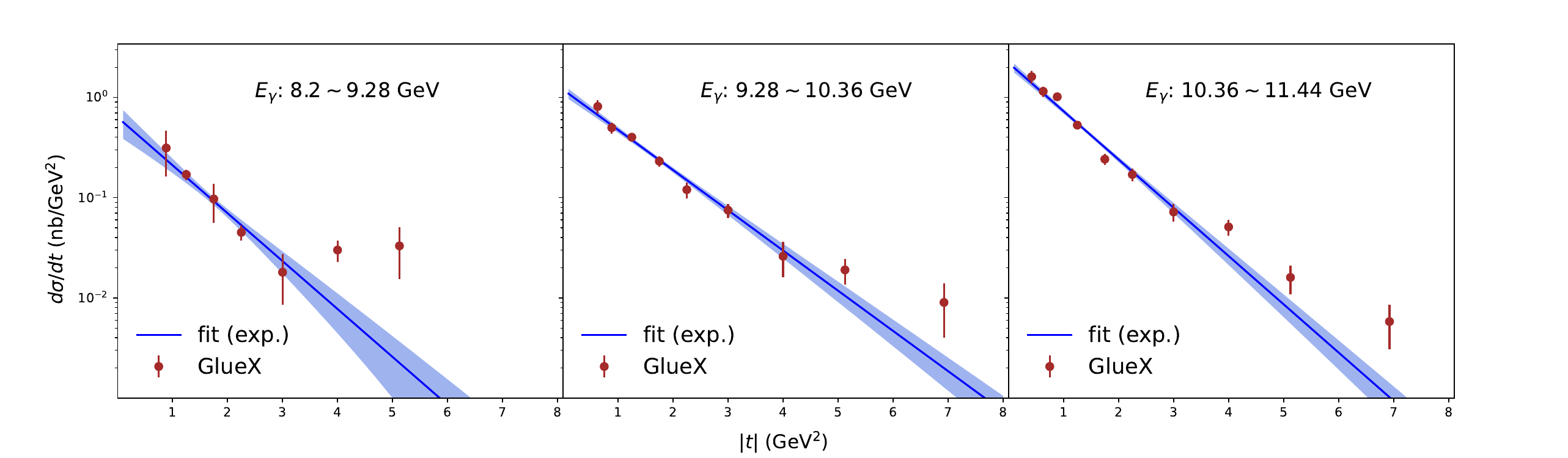}}
	\subfigure{
		\includegraphics[width=0.9\textwidth]{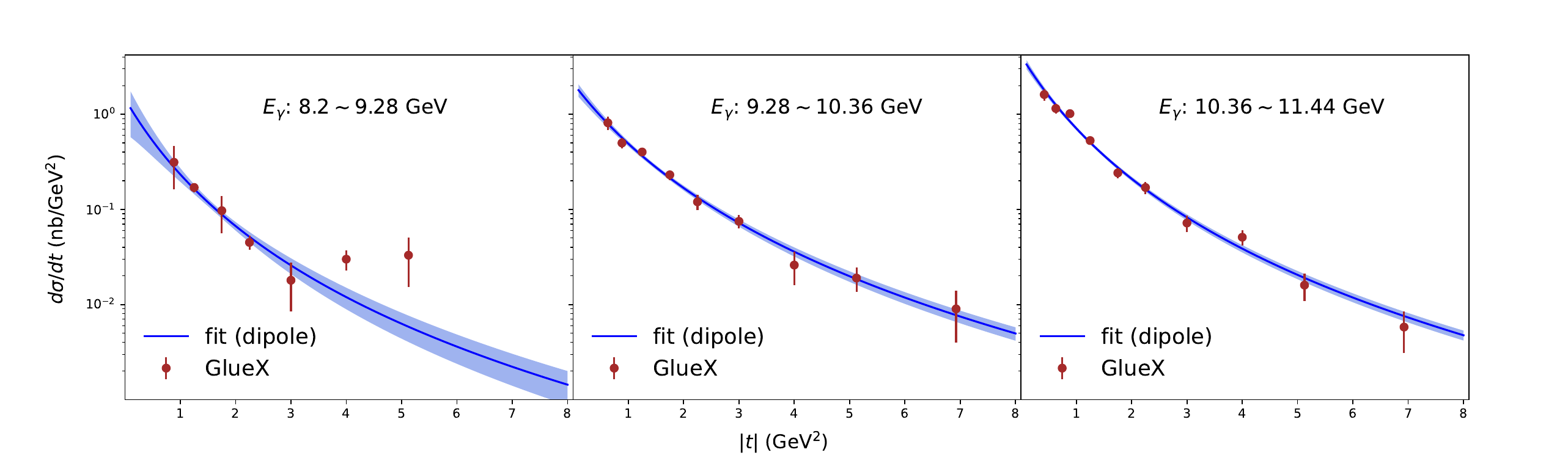}}
	\caption{Exponential (up) and dipole (down) fits to the differential cross section $d\sigma/dt$ using the data from GlueX collaboration \cite{GlueX:2023pev}. The light blue band shows its uncertainty. The parameters are listed in Tables \ref{tab:gluex} and \ref{tab:gluex-dipole}.}
	\label{fig:fit_GlueX}
\end{figure*}

%\noindent\emph{3.$\;$Numerical results and discussions} ---
The exponential and dipole fits to the differential cross-section $d\sigma/dt$ utilizing data from the $J/\psi-007$ and GlueX collaborations are presented in Figures \ref{fig:fit_hall_c} to \ref{fig:fit_GlueX}. The corresponding fitting parameters are detailed in Tables \ref{tab:jpsi007} to \ref{tab:gluex-dipole}. To extract the proton trace anomaly and sigma term, we adopt the charm quark mass with $r_0$ in Eq. (\ref{eq:FXN}) as $m_c=1.4$ GeV \cite{Wang:2019mza,Cui:2019dwv,Watt:2007nr}. The strong coupling $\alpha_s$ is cited from \cite{Deur:2016tte,Deur:2023dzc} at leading order
\begin{equation}
	\alpha_\mathrm{s}\left(\mu^2\right)=\frac{4\pi}{\beta_0\ln\left[\left(m_\alpha^2+\mu^2\right)/\Lambda^2\right]},
	\label{eq:as}
\end{equation}
where $\Lambda=0.58$ GeV is the landau pole \cite{Deur:2016tte,Deur:2023dzc} with momentum-subtraction \cite{Zafeiropoulos:2019flq} and $\beta_0=(33-2n_f)/3$ with quark flavor number $n_f=4$. The parameter $m_\alpha$ represents the gluon mass scale \cite{Cui:2019dwv}. In the next subsection of this section, we will discuss the impact of estimating the running strong coupling constant on the results.


\begin{table}[htbp]
					\caption{Exponential fit to the differential cross section $d\sigma/dt$ using the data from $J/\psi-007$ collaboration \cite{Duran:2022xag}. Photon energy, fitting parameters, trace anomaly, strangeness sigma term and reduced $\chi^2$.}
	\resizebox{\columnwidth}{!}{%
		\begin{tabular}{c|c|c|c|c|c}
			\hline
			$E_\gamma$ (GeV) & $\frac{d\sigma}{dt}|_{t\to0}$ ($\frac{\mathrm{nb}}{\mathrm{GeV^2}}$) & $B$ (GeV$^{-2}$) & $M_a/M_N$     & $\sigma_{sN}$ (MeV) & $\chi_\mathrm{red.}^2$\\ \hline
			$9.1\sim9.25$    & $0.57\pm0.23$    & $0.90\pm0.30$    & $0.08\pm0.02$ & $543.75\pm66.83$    &0.4 \\ \hline
			$9.25\sim9.4$    & $0.42\pm0.10$    & $0.57\pm0.17$    & $0.08\pm0.01$ & $596.21\pm33.19$    &0.9\\ \hline
			$9.4\sim9.55$    & $0.96\pm0.19$    & $1.00\pm0.18$    & $0.11\pm0.01$ & $463.73\pm40.21$   & 0.7\\ \hline
			$9.55\sim9.7$    & $1.20\pm0.23$    & $1.06\pm0.20$    & $0.12\pm0.01$ & $443.81\pm44.27$   &0.4 \\ \hline
			$9.7\sim9.85$    & $1.82\pm0.49$    & $1.52\pm0.25$    & $0.14\pm0.02$ & $343.38\pm71.63$  &  0.5\\ \hline
			$9.85\sim10.0$   & $1.54\pm0.41$    & $1.31\pm0.24$    & $0.13\pm0.02$ & $402.78\pm63.78$   & 0.9\\ \hline
			$10.0\sim10.15$  & $1.41\pm0.25$    & $1.05\pm0.15$    & $0.12\pm0.01$ & $440.65\pm39.29$   & 1.0\\ \hline
			$10.15\sim10.3$  & $1.67\pm0.26$    & $1.17\pm0.15$    & $0.12\pm0.01$ & $420.65\pm35.16$   & 1.5\\ \hline
			$10.3\sim10.45$  & $1.63\pm0.39$    & $1.28\pm0.29$    & $0.12\pm0.01$ & $443.99\pm52.05$   &0.7 \\ \hline
			$10.45\sim10.6$  & $1.97\pm0.47$    & $1.39\pm0.30$    & $0.12\pm0.01$ & $420.93\pm54.28$  &  0.7\\ \hline
		\end{tabular}%
	}
	\label{tab:jpsi007}
\end{table}

\begin{table}[htbp]
	\caption{Dipole fit to the differential cross section $d\sigma/dt$ using the data from $J/\psi-007$ collaboration \cite{Duran:2022xag}. Photon energy, fitting parameters, trace anomaly, strangeness sigma term and reduced $\chi^2$.}
	\resizebox{\columnwidth}{!}{%
		\begin{tabular}{c|c|c|c|c|c}
			\hline
			$E_\gamma$ (GeV) & $\frac{d\sigma}{dt}|_{t\to0}$ ($\frac{\mathrm{nb}}{\mathrm{GeV^2}}$)  & $M_A$ (GeV) & $M_a/M_N$     & $\sigma_{sN}$ (MeV) & $\chi_\mathrm{red.}^2$\\ \hline
			$9.1\sim9.25$    & $0.77\pm0.48$    & $1.68\pm0.44$    & $0.10\pm0.03$ & $487.85\pm116.9$    &0.44 \\ \hline
			$9.25\sim9.4$    & $0.47\pm0.15$    & $2.30\pm0.48$    & $0.08\pm0.01$ & $580.32\pm44.64$    &0.96\\ \hline
			$9.4\sim9.55$    & $1.30\pm0.41$    & $1.59\pm0.22$    & $0.13\pm0.02$ & $395.48\pm68.70$   & 0.90\\ \hline
			$9.55\sim9.7$    & $1.73\pm0.50$    & $1.51\pm0.17$    & $0.14\pm0.02$ & $339.43\pm73.45$   &0.39 \\ \hline
			$9.7\sim9.85$    & $3.35\pm1.52$    & $1.17\pm0.16$    & $0.19\pm0.04$ & $152.36\pm158.6$  &  0.49\\ \hline
			$9.85\sim10.0$   & $2.75\pm1.15$    & $1.26\pm0.17$    & $0.17\pm0.04$ & $242.79\pm131.5$   & 0.86\\ \hline
			$10.0\sim10.15$  & $2.22\pm0.64$    & $1.43\pm0.16$    & $0.15\pm0.02$ & $329.19\pm76.84$   & 1.21\\ \hline
			$10.15\sim10.3$  & $2.83\pm0.80$    & $1.31\pm0.14$    & $0.16\pm0.02$ & $281.52\pm73.86$   & 2.03\\ \hline
			$10.3\sim10.45$  & $2.53\pm0.87$    & $1.32\pm0.19$    & $0.14\pm0.04$ & $337.13\pm101.3$   &0.48 \\ \hline
			$10.45\sim10.6$  & $2.79\pm0.99$    & $1.32\pm0.20$    & $0.15\pm0.02$ & $334.38\pm90.59$  &  0.74\\ \hline
		\end{tabular}%
	}
	\label{tab:jpsi007-dipole}
\end{table}

%\begin{table}[htbp]
%		\caption{Exponential fit to the differential cross section $d\sigma/dt$ using the data from $J/\psi-007$ collaboration \cite{Deur:2023dzc}. Fitting parameters, trace anomaly and strangeness sigma term.}
%	\begin{tabular}{c|c|c|c|c}
%		\hline
%		$E_\gamma$ (GeV) & $A$ (nb/GeV$^2$) & $B$ (GeV$^{-2}$) & $M_a/M_N$     & $\sigma_{sN}$ (MeV) \\ \hline
%		$9.1\sim9.25$    & $0.57\pm0.23$    & $0.90\pm0.30$    & $0.10\pm0.02$ & $493.54\pm76.81$    \\ \hline
%		$9.25\sim9.4$    & $0.42\pm0.10$    & $0.57\pm0.17$    & $0.08\pm0.01$ & $563.89\pm36.94$    \\ \hline
%		$9.4\sim9.55$    & $0.96\pm0.19$    & $1.00\pm0.18$    & $0.12\pm0.01$ & $427.03\pm43.73$    \\ \hline
%		$9.55\sim9.7$    & $1.20\pm0.23$    & $1.06\pm0.20$    & $0.12\pm0.01$ & $413.24\pm47.35$    \\ \hline
%		$9.7\sim9.85$    & $1.82\pm0.49$    & $1.52\pm0.25$    & $0.15\pm0.02$ & $311.95\pm75.82$    \\ \hline
%		$9.85\sim10.0$   & $1.54\pm0.41$    & $1.31\pm0.24$    & $0.13\pm0.02$ & $377.74\pm67.12$    \\ \hline
%		$10.0\sim10.15$  & $1.41\pm0.25$    & $1.05\pm0.15$    & $0.12\pm0.01$ & $418.21\pm41.29$    \\ \hline
%		$10.15\sim10.3$  & $1.67\pm0.26$    & $1.17\pm0.15$    & $0.13\pm0.01$ & $395.97\pm37.04$    \\ \hline
%		$10.3\sim10.45$  & $1.63\pm0.39$    & $1.28\pm0.29$    & $0.12\pm0.01$ & $417.89\pm55.16$    \\ \hline
%		$10.45\sim10.6$  & $1.97\pm0.47$    & $1.39\pm0.30$    & $0.13\pm0.02$ & $389.26\pm58.02$    \\ \hline
%	\end{tabular}
%	\label{tab:jpsi007}
%\end{table}

\begin{table}[htbp]
	\caption{Exponential fit to the differential cross section $d\sigma/dt$ using the data from GlueX collaboration \cite{GlueX:2023pev}. Photon energy, fitting parameters, trace anomaly, strangeness sigma term and reduced $\chi^2$.}
	\resizebox{\columnwidth}{!}{%
		\begin{tabular}{c|c|c|c|c|c}
			\hline
			$E_\gamma$ (GeV) & $\frac{d\sigma}{dt}|_{t\to0}$  ($\frac{\mathrm{nb}}{\mathrm{GeV^2}}$)             & $B$ (GeV$^{-2}$) & $M_a/M_N$     & $\sigma_{sN}$ (MeV) & $\chi_\mathrm{red.}^2$\\ \hline
			$8.92\sim9.28$   & $0.63\pm0.21$                & $1.10\pm0.19$    & $0.10\pm0.02$ & $516.52\pm61.11$   & 2.9\\ \hline
			$9.28\sim10.36$  & $1.20\pm0.15$ & $0.92\pm0.07$    & $0.11\pm0.01$ & $453.22\pm26.39$   &1.2 \\ \hline
			$10.36\sim11.44$ & $2.21\pm0.24$                & $1.11\pm0.07$    & $0.12\pm0.01$ & $433.67\pm24.66$   & 3.2\\ \hline
		\end{tabular}%
	}
	\label{tab:gluex}
\end{table}

\begin{table}[htbp]
	\caption{Dipole fit to the differential cross section $d\sigma/dt$ using the data from GlueX collaboration \cite{GlueX:2023pev}. Photon energy, fitting parameters, trace anomaly, strangeness sigma term and reduced $\chi^2$.}
	\resizebox{\columnwidth}{!}{%
		\begin{tabular}{c|c|c|c|c|c}
			\hline
			$E_\gamma$ (GeV) & $\frac{d\sigma}{dt}|_{t\to0}$ ($\frac{\mathrm{nb}}{\mathrm{GeV^2}}$)  & $M_A$ (GeV) & $M_a/M_N$     & $\sigma_{sN}$ (MeV) & $\chi_\mathrm{red.}^2$\\ \hline
			$8.92\sim9.28$   & $1.45\pm0.82$                & $1.31\pm0.19$    & $0.14\pm0.04$ & $330.24\pm154.9$   & 2.0\\ \hline
			$9.28\sim10.36$  & $2.13\pm0.35$                & $1.50\pm0.07$    & $0.15\pm0.01$ & $310.82\pm41.86$   &0.6 \\ \hline
			$10.36\sim11.44$ & $4.14\pm0.50$                & $1.34\pm0.05$    & $0.16\pm0.01$ & $269.02\pm37.13$   & 1.1\\ \hline
		\end{tabular}%
	}
	\label{tab:gluex-dipole}
\end{table}

%\begin{table}[htbp]
%	\caption{Exponential fit to the differential cross section $d\sigma/dt$ using the data from GlueX collaboration \cite{GlueX:2023pev}. Fitting parameters, trace anomaly and strangeness sigma term.}
%		\begin{tabular}{c|c|c|c|c}
%			\hline
%			$E_\gamma$ (GeV) & $A$ (nb/GeV$^2$)             & $B$ (GeV$^{-2}$) & $M_a/M_N$     & $\sigma_{sN}$ (MeV) \\ \hline
%			$8.92\sim9.28$   & $0.63\pm0.21$                & $1.10\pm0.19$    & $0.12\pm0.02$ & $431.19\pm75.41$    \\ \hline
%			$9.28\sim10.36$  & $1.20\pm0.15$ & $0.92\pm0.07$    & $0.12\pm0.01$ & $429.97\pm27.81$    \\ \hline
%			$10.36\sim11.44$ & $2.21\pm0.24$                & $1.11\pm0.07$    & $0.13\pm0.01$ & $391.58\pm26.95$    \\ \hline
%		\end{tabular}%
%	\label{tab:gluex}
%\end{table}

\begin{figure*}[htbp]
	\centering  %图片全局居中
	\subfigure{
		\includegraphics[width=0.45\textwidth]{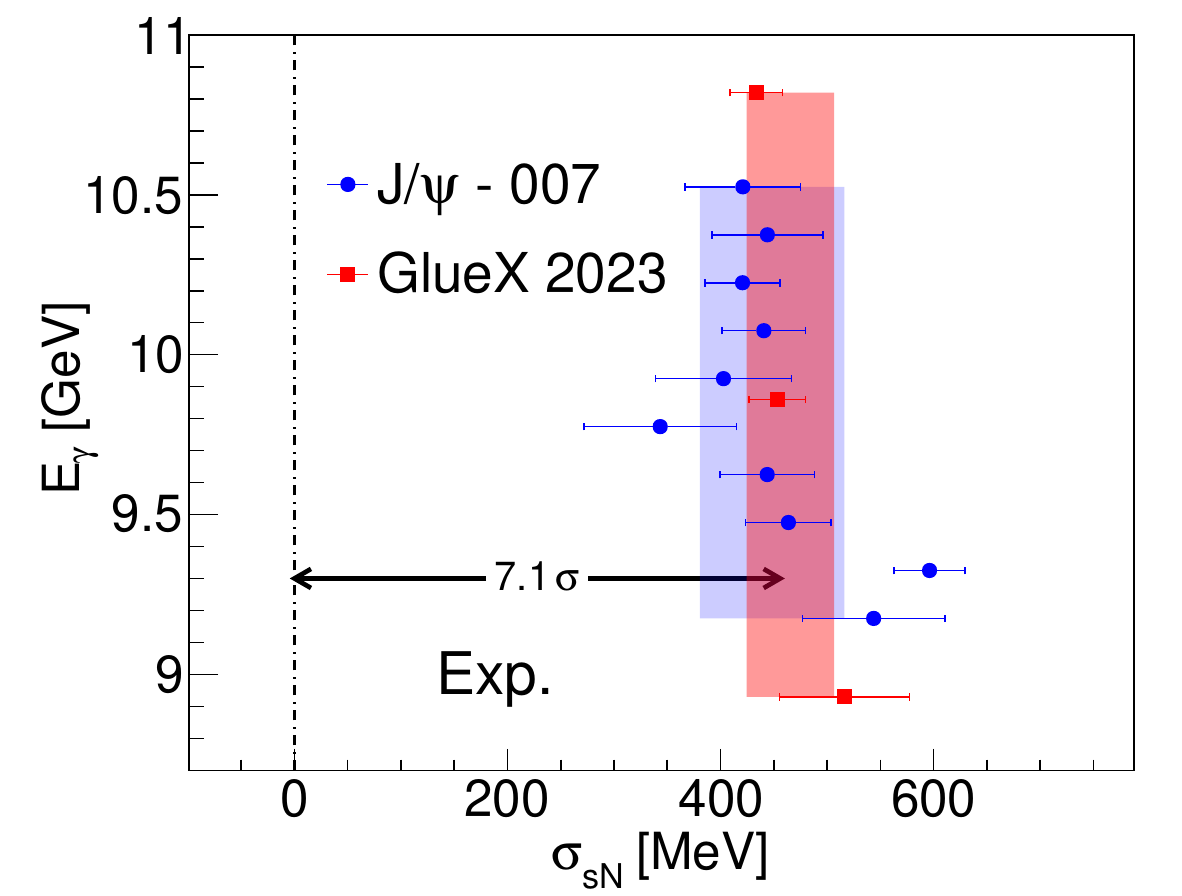}}
	\subfigure{
		\includegraphics[width=0.45\textwidth]{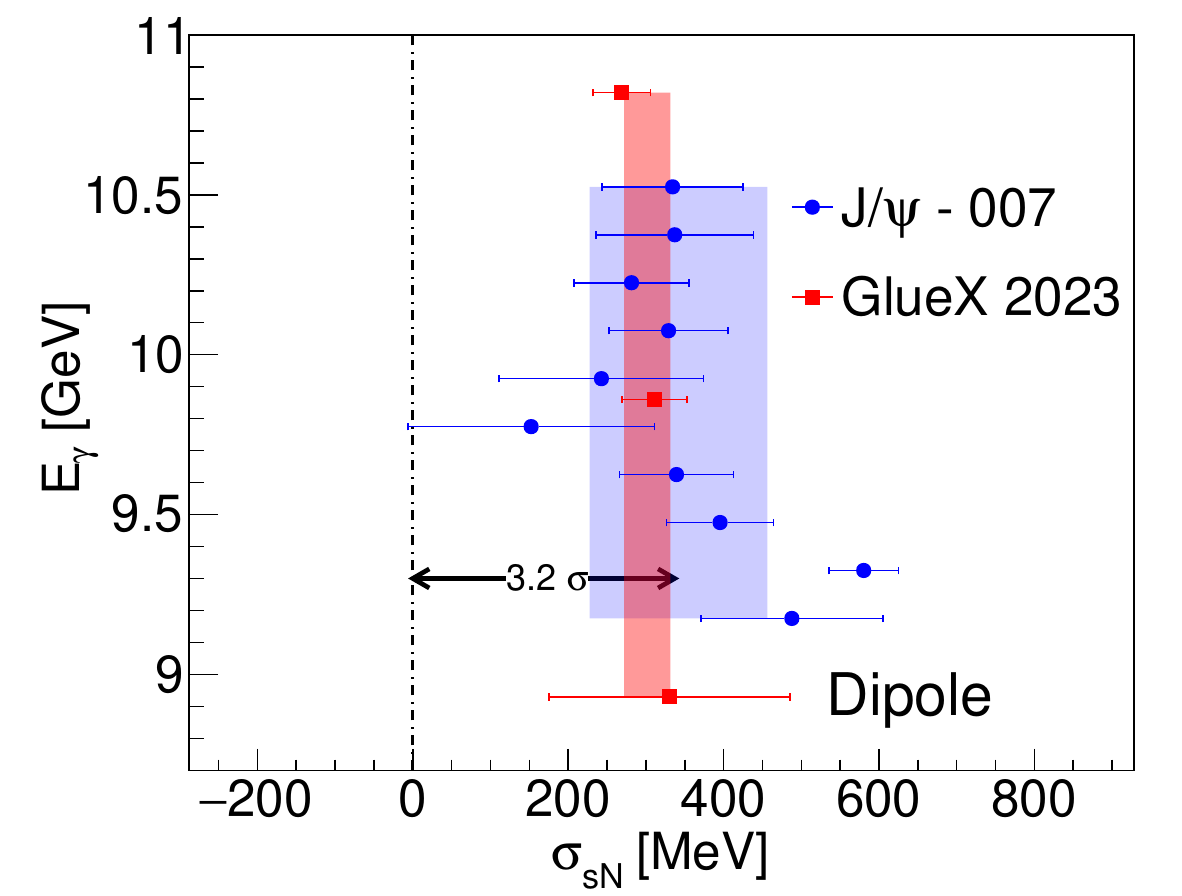}}
	\caption{The extracted strangeness sigma terms of proton with the experiments data from $J/\psi-007$ (blue points) and GlueX (red squares) collaborations \cite{Duran:2022xag,GlueX:2023pev} with exponential (left) and dipole (right) fits. The blue and red shades present the corresponding standard deviations of data sets. The difference between the mean of all extractions and zero are $7.1$ and $3.2$ times combined standard deviations corresponding two types of fits, respectively.}
	\label{fig:simgaterm}
\end{figure*}

Figure \ref{fig:simgaterm} illustrates the extracted strangeness sigma terms of the proton using data sets from the $J/\psi-007$ (blue circles) and GlueX (red squares) collaborations \cite{Duran:2022xag,GlueX:2023pev} with two types fit discussed above. The shaded regions of the corresponding colors indicate the standard deviations of each data set at different energy settings. The combined standard deviation of the two data sets are $\sigma=64.6$ MeV (exponential fit) and $106.8$ MeV (dipole fit). Here, the calculated standard deviation only considers the standard deviation of the extracted strangeness sigma term itself as a function of energy, without taking into account the systematic uncertainties introduced by the model and parameters. The impact of systematic uncertainties on the results is discussed in next subsection. Based on these results, we infer that the strangeness sigma term for the proton has statistical significances of $7.1\sigma$ (exponential fit) and $3.2 \sigma$ (dipole fit),  suggesting that it is not zero.

Furthermore, we conduct the Kolmogorov-Smirnov (KS) test \cite{Eadie:100342} on the sigma term distributions extracted from the two sets of experimental data. Initially, we test the difference between the distributions of the two sets of sigma terms and the ZERO values, yielding a p-value of $0.0$ for both, we note the extraction results from both fits support this conclusion. Additionally, we test the similarity of the two distributions themselves and obtained the p-values of approximately 0.5 with both fits, indicating that the two sets of experimental data cannot be dismissed as being the same distribution. It is important to note that, due to the relatively small number of sample points, the results of the KS test serve solely as a reference and do not definitively determine the conclusions we propose regarding the KS test.

We proceed to discuss the disparities between our approach and the previous works \cite{Wang:2019mza,Wang:2022tzw}. Firstly, the dimensional analysis of the $J/\psi-N$ scattering amplitude in Eq. (\ref{eq:FXN}) differs from that of the previous works. We have modified it in accordance with the approach detailed in Ref. \cite{Kharzeev:1998bz}. Secondly, we have developed a new running coupling constant energy scale tailored to the different experimental data energy settings. Specifically, we utilize the difference between the energy of the center-of-mass system of the photon-proton and the mass of the final-state proton and $J/\psi$ as the $\alpha_s(\mu)$ energy scale, i.e., $\mu=W-M_N-m_{J/\psi}$. By introducing a new definition for $\mu$, the scale we provide corresponds to approximately $\mu^2=0.025\sim 0.32$ GeV$^2$ under the experimental conditions of two experiments, with the corresponding coupling constant being $\alpha_s=1.14 \sim 1.35$ with Eq. (\ref{eq:as}). Here the $\alpha_s$'s correspond to the values at each energy scales and do not represent the uncertainties them introduce. This approach simplifies the photon coupling to vector mesons as described by the VMD approach. However, in the non-relativistic approximation, photon energies may impact the binding of charm quark pairs into $J/\psi$, subsequently affecting the Bohr radius $r_0$. The center-of-mass energy of the system can be derived from the photon energy as input, and the energy remaining after removing the mass of the particles produced in the final state can be roughly understood as the binding energy required for the quasi-meson to meson process.

Lastly, our extracted sigma term values are larger than those presented in other works (see Fig. 18 in Ref. \cite{Alexandrou:2019brg}). For example, the chiral perturbation theory calculation yields a result of approximately $\sigma_{sN}=16(80)$ MeV \cite{Alarcon:2012nr}, while the lattice calculation gives $\sigma_{sN}=40(12)$ \cite{Yang:2015uis} and the more precise lattice result is $\sigma_{sN}=45.6(6.2)$ \cite{Alexandrou:2019brg}. While larger uncertainties in the experimental extraction cannot be disregarded, we have assumed the contribution of only three flavor quarks inside the proton throughout the extracting process, based on which we obtain a strangeness sigma term of about 400 MeV. Building on these assumptions, we obtain a relatively centralized distributions of the proton trace anomaly and strangeness sigma term. It is important to emphasize that our results depend on the validity of the VMD model and the estimation of the model parameters. The relatively concentrated distribution of the strangeness sigma term with respect to the photon energy inputs implies a weak dependence of the sigma term on the experiments energies.

\subsection{Uncertainties analysis}
Now, let us briefly discuss the impact of the VMD model's validity on extracting the strangeness sigma term. The frequent use of the VMD model in connecting quarkonium photo-production processes and the elastic scattering amplitudes of quarkonium and protons, such as in the extraction of vector meson-nucleon (nuclear) scattering lengths \cite{Strakovsky:2014wja,Strakovsky:2019bev,Strakovsky:2020uqs,Pentchev:2020kao,Wang:2022xpw,Wang:2022zwz,Han:2022khg,Han:2022btd}, is noteworthy. However, discussions on the validity of the VMD model and the uncertainties in its relevant model parameters remain scarce. Recently, there has been a discussion on the partial wave analysis of the near-threshold photo-production process of $J/\psi$ \cite{JointPhysicsAnalysisCenter:2023qgg}. The authors argue that the estimation of the $J/\psi-p$ elastic scattering amplitude using the VMD model is debatable. They have considered the partial wave amplitudes for the corresponding process and have defined the ratio of the VMD model amplitude to the partial wave amplitude $R_\mathrm{VMD}=|F_{\mathrm{VMD}}/F_\mathrm{PW}|$ to assess the suitability of the model.
In principle, if the VMD model is exact, the experimentally driven partial wave analysis would result in $R_\mathrm{VMD}=1$. Please refer to the relevant section in Ref. \cite{JointPhysicsAnalysisCenter:2023qgg} for the estimation and analysis of $R_\mathrm{VMD}$. The authors argue that the VMD model's estimated elastic scattering amplitude is two orders of magnitude smaller than the partial wave result. This would lead to significant biases when using VMD to extract scattering lengths or other observables related to elastic scattering amplitudes. Additionally, the amplitude estimation technique used in this letter, which depends on the parameter $b$, introduces uncertainties arising from the quarkonium Bohr radius and coupling constant. The resulting quantum anomalous energy of the proton and the extraction of the strangeness sigma term will be affected by these uncertainties. For a discussion on this matter, please refer to our previous work on the analysis of parameter uncertainties \cite{Kou:2021bez}. 

In fact, the core issue of the entire extraction work is to determine the parameter $r_0$ in Eq. (\ref{eq:FXN}), which corresponds to the charm quark mass and the strong coupling constant $\alpha_s$. On the one hand, the charm quark mass is always subject to different choices in handling various problems in particle physics, ranging from $m_c=1.2$ GeV to $m_c=1.67$ GeV \cite{Workman:2022ynf}. We utilize the corresponding charm quark mass $m_c=1.4$ GeV as input, because it was used to calculate the $J/\psi$ photo-productions \cite{Watt:2007nr}. In principle, it is valuable to analyze the fluctuations of the quark mass and their corresponding impact on the strangeness sigma term. Similar discussions can be found in Ref. \cite{Kou:2021bez}. We also performed a similar analysis to estimate the overall uncertainty introduced by fluctuations in parameter values. We found that when the fluctuations in quark mass are approximately 5$\%$, it ultimately affects the extracted average value of the strangeness sigma term by about 12$\%$, and the statistical significance fluctuates by approximately 2 times of standard deviations. Meanwhile, the fluctuations in the running coupling constant have a similar impact on the analysis as the fluctuations in quark mass. Likewise, the selection of the coupling constant that describes the binding of charm and anti-charm quarks is also a non-perturbative issue. However, when we need a specific value for the coupling constant as a reference, some method is still required to analyze the behavior of the coupling constant in the infrared region, even though it often diverges within the framework of perturbative QCD. The infrared saturation of $\alpha_s$ has advantages in describing phenomena such as low-energy hadron structure \cite{Deur:2016tte,Deur:2023dzc}. Another important factor that influences our extraction of the strangeness sigma term is the forward scattering information obtained by fitting the differential cross-section to obtain the zero momentum transfer. In Ref. \cite{Duran:2022xag}, the analysis of the exponential and dipole form factor parameterizations was conducted, and the uncertainty arising from the extrapolation of the fit to the unphysical region ($|t|=0$) was examined. This uncertainty depends on the parameterization form itself and ultimately manifests in the extraction of the sigma term. In the paper, we also conducted a differential test of different parameterizations, and the results are similar to Ref. \cite{Duran:2022xag}. Finally, we argue that the precise analysis of the VMD model and its parameter uncertainties relies on constraints from high-precision experiments.
\smallskip

\section{Conclusion and outlook}
\label{sec:Conclusion}
%\noindent\emph{4.$\;$Conclusions} ---
In conclusion, we have successfully extracted the strangeness sigma term for the proton from experimental data for the first time, establishing its photon energies independence. This constitutes the first finding of a non-zero contribution to the intrinsic strangeness of the proton being quantified in terms of the proton mass contribution. The strangeness sigma term is determined to be $455.62\pm64.60$ MeV (exponential fit) and $337.74\pm106.79$ MeV (dipole fit), demonstrating a statistical significance of at least 7.1 (exponential fit) and 3.2 (dipole fit) times the standard deviation between its value and zero. From our perspective, the strangeness component is not negligible, at least in terms of its contribution to the proton mass, as evidenced by the extraction of $\sigma_{sN}$. Additionally, it is important to emphasize that the current extraction methods are model-dependent, and the systematic uncertainties are significant, requiring more relevant experiments to constrain the model parameters. However, there are substantial uncertainties in the available experimental data, including those introduced by the photon energy bins. Therefore, the ongoing construction of the Electron-Ion Collider in the United States  \cite{Accardi:2012qut,AbdulKhalek:2021gbh} and China \cite{Chen:2018wyz,Chen:2020ijn,Anderle:2021wcy} will be necessary and timely for improving experimental precision.

\begin{acknowledgments}
We would like to express our gratitude for the discussions with Dr. Rong Wang, and also acknowledge Dr. Daniele Anderle for providing valuable insights related to the partial wave analysis. We also appreciate the suggestions and comments from the reviewers during the peer review process, which helped us in revising the manuscript. This work is supported by the Strategic Priority Research Program of Chinese Academy of Sciences (Grant NO. XDB34030301).
\end{acknowledgments}

\bibliographystyle{apsrev4-1}
\bibliography{refs}
\newpage

\end{document}